\documentclass[fleqn]{elsarticle}
\usepackage{graphicx}
\usepackage{amsmath,amssymb,amsfonts,bm}

\title{Theoretical and computational study of the energy dependence of
  the muon transfer rate from hydrogen to higher-Z gases}

 \author[inrne]{Dimitar Bakalov\corref{cor1}}
 \author[ifj]{Andrzej Adamczak}
 \author[inrne]{Mihail Stoilov}
 \ead[url]{dbakalov@inrne.bas.bg}
 \author[infn2]{Andrea Vacchi}
 \cortext[cor1]{Corresponding author}
 \address[inrne]{Institute for Nuclear Research and Nuclear Energy,
 Bulgarian Academy of sciences, Tsarigradsko chauss\'{e}e 72, Sofia
 1784, Bulgaria}
 \address[ifj]{Institute of Nuclear Physics, Polish Academy of Sciences,
ul. Radzikowskiego 152, 31-342 Krakow, Poland}
 \address[infn2]{Istituto Nazionale di Fisica Nucleare, Sezzione di
 Trieste, Via A. Valerio 2, 34127 Trieste, Italy}

\begin{document}

\begin{abstract}
  The recent PSI Lamb shift experiment and the controversy about proton
  size revived the interest in measuring the hyperfine
  splitting in muonic hydrogen as an alternative possibility for
  comparing ordinary and muonic hydrogen spectroscopy data on proton
  electromagnetic structure.
  This measurement critically depends on the energy dependence of the muon
  transfer rate to heavier gases in the epithermal range. The available
  data provide only qualitative information, and the theoretical
  predictions have not been verified.
  We propose a new method by measurements of the transfer rate in thermalized
  target at different temperatures, estimate its accuracy and
  investigate the optimal experimental conditions.
\end{abstract}

\maketitle

 \section{Introduction}

 The laser spectroscopy measurement of the hyperfine structure of the
 ground state of the muonic hydrogen atom was first proposed more than
 two decades ago \cite{pla93}, motivated by the understanding that in
 some sense it would be complementary to the top accuracy measurements
 of the hyperfine splitting (HFS) in ordinary hydrogen \cite{essen}.
 Subsequent studies revealed, however, some severe difficulties to this
 goal: the absence of a tunable near-infrared (NIR) laser with
 sufficient power that necessitates the use of a multi-pass optical
 cavity to enhance the stimulation of transitions between the hyperfine
 levels, which in turn makes inapplicable the initially proposed
 experimental method. In the following years an alternative experimental
 method was proposed that exploits specific features of the muon
 transfer reaction at epithermal collision energies \cite{hfi2001} and
 is compatible with the use of a multi-pass cavity. Also, the recent
 progress in the development of pulsed NIR sources of monochromatic
 radiation and IR optics \cite{lasers} has brought the produced laser
 power close to the needed magnitudes. All this made the experimental
 project look feasible, though still remaining a significant challenge.
 What supplied the missing motivation to start the work on the project
 were the results of the recent muonic hydrogen Lamb shift experiment at
 PSI \cite{PSI} and the discovered incompatibility of the values of the
 charge radius of the proton extracted from muonic hydrogen spectroscopy
 on the one hand, and ordinary hydrogen spectroscopy and electron-proton
 scattering data, on the other. The few years of intense search for an
 explanation of this discrepancy have not lead to any conclusion, and
 the most direct way to throw light on the subject --- next to an
 independent new measurement of the proton charge radius --- appears to
 be the measurement of the hyperfine splitting in muonic hydrogen and
 extracting from it the proton Zemach radius: confirming the value
 obtained from hydrogen spectroscopy would probably give more weight to
 the explanations supposing methodology uncertainties in the
 hydrogen-based proton charge radius, while in the opposite case the
 search for new physics will have good reasons to be intensified.

 With this motivation in mind, as a first stage of the preparation for
 the measurement of the HFS in muonic hydrogen,
 we started the thorough study of the efficiency and
 accuracy of the method proposed for this experiment in
 Ref.~\cite{hfi2001}. Its physical background is simple: the muonic
 hydrogen atoms that absorb a photon of the resonance frequency, undergo
 a hyperfine para-to-ortho transition. When de-excited back to the
 para spin state in a collision with a H$_2$ molecule, these atoms are accelerated
 by $\sim{}0.1$~eV \cite{hfi2001}. The number of atoms that have
 undergone the above sequence (and therefore --- the tuning of the laser
 to the resonance hyperfine transition frequency) can be determined by
 observing a reaction whose rate depends on the energy of the $\mu^-p$
 atoms. In Ref.~\cite{hfi2001} it was proposed to observe the transfer
 of the muon from muonic hydrogen to the nucleus of a heavier gas added
 to the hydrogen target for which the transfer rate is energy-dependent
 in the epithermal range. There are experimental evidences that for some
 gases (e.g., oxygen, argon, etc.)  this is indeed the case
 \cite{oxygen2,neon2,argon2}. The goal of our investigations is to
 obtain reliable quantitative data on the energy dependence of the rate
 of muon transfer to these gases --- and possibly other as well --- in
 order to select the optimal chemical composition and physical
 parameters (pressure, temperature) of the hydrogen gas target that will
 provide the highest accuracy in the future measurements of the
 hyperfine splitting of $(\mu^-p)_{1s}$ and the Zemach radius of the
 proton.  In the present paper we focus at the main physical processes
 with $\mu^-p$ atoms in a mixture of hydrogen and a higher-Z gas,
 and deduce the optimal experimental
 conditions for the measurement of the energy dependence of the muon
 transfer rate to the admixture gas. The results have been obtained with a
 Monte Carlo simulation code that uses the cross sections and rates for
 the scattering $\mu^-p+\mathrm{H}_2$, which were calculated in
 Refs.~\cite{adam96,adam06}. The code has been verified on the many
 preceding occasions (see, e.g.,~\cite{argon2,werth,adam07}).

 \section{Preceding results and novelty of the proposed measurements}

 The muon transfer in collisions of muonic hydrogen atoms with atoms of
 higher-Z gas admixtures has been actively investigated for many years
 because of the important impact of hydrogen impurities on the
 observable rates of specific reactions such as nuclear muon capture
 \cite{mucap}, muonic molecule formation \cite{faifman}, etc. In the
 limit of low collision energies, theory predicts in the lowest order
 approximation a flat energy dependence of the muon transfer rate
 \cite{landau,gerstein}, while more refined calculations show that for
 some gases its value may vary by up to an order of magnitude for
 collision energies below 1~eV \cite{dupays}. Some experimental data,
 e.g., on transfer to sulfur \cite{oxygen2}, consist of a single
 data point,
 the average rate of muon transfer at certain pressure and
 temperature, obtained as the disappearance rate of muonic hydrogen
 atoms and determined from the time distribution of the events of
 de-excitation of the muonic atoms of the heavier gas. In the case of
 oxygen \cite{oxygen2}, argon \cite{argon2} and neon \cite{neon2},
 however, the time distributions of these events clearly display two
 different time ranges with substantially different disappearance
 exponents for the hot and the thermalized atoms.  This was interpreted
 as an evidence for a non-flat energy dependence of the muon transfer
 rate \cite{werth}, and a step-function was introduced to describe
 qualitatively the energy dependence of the transfer rate to oxygen,
 argon and neon.  The earlier estimates of the efficiency of the
 experimental method in \cite{pla93,hfi2001} were obtained using this
 step-function approximation. Its accuracy, however, is far from what is
 needed for the planning and optimization of the muonic hydrogen
 hyperfine experiment because the contribution from the atoms with
 energies in the whole epithermal range is averaged over their energy
 distribution and is accounted for with a single parameter --- the
 disappearance slope for the ``unexpected delayed''
 events~\cite{oxygen2} --- which incorporates the uncertainties due to
 the strongly model-dependent energy distribution of the epithermal
 atoms \cite{jens02,pohl06,faif08}.  To solve the problem, we propose to
 perform instead a series of measurements of the muon transfer rate from
 {\em thermalized muonic hydrogen atoms} at different temperatures in an
 as broad as possible temperature range. The numerical technique of
 investigating the energy dependence of the muon transfer rate, which is
 presented in Sect.~\ref{sect:rate}, relies substantially on the
 assumption that the muonic hydrogen atoms are thermalized and their
 energy distribution --- the Maxwell-Boltzmann distribution --- contains
 no uncertainties.  Our study in Sect.~\ref{sect:processes} is also
 focused on the processes in a mixture of hydrogen and higher-Z gases at
 thermal equilibrium.

 \section{Processes involving the muonic hydrogen atoms in a mixture of
   hydrogen and higher-Z gases}
 \label{sect:processes}

 When slowed down and stopped in a mixture of hydrogen and higher-Z
 gases, a negative muon is captured by the Coulomb field of the nuclei
 and forms exotic muonic atoms in which one (or more) of the electrons
 are ejected and replaced by the muon. We do not consider here the
 processes of muon capture in an excited state of the exotic atom and
 the subsequent de-excitation by competing Auger and radiative
 transitions and in non-elastic collisions because for the hydrogen
 target densities of interest (pressures above 10~atm at room
 temperature) the whole cascade down to the ground $1s$ state takes no
 more than about 1~ns. This choice of the target density is determined
 by the specific features of the available gas containers: simulations
 (to be reported elsewhere) with the FLUKA code~\cite{fluka} have shown
 that too few muons are stopped in the target at lower densities. So, to
 a good approximation, the description of the history of a muonic atom
 can start from the ground $1s$ state.

 In the general case (when the target gas is a natural mixture of
 hydrogen and deuterium), the following processes take place: elastic
 scattering of the muonic atoms, spin-flip, muon exchange between the
 hydrogen isotopes, and rotational-vibrational transitions in target
 molecules; formation of the $pp\mu$, $pd\mu$, and $dd\mu$ molecules;
 nuclear fusion in $pd\mu$ and $dd\mu$; nuclear muon capture; muon
 decay, and muon transfer from muonic hydrogen atoms to atomic states of
 higher-Z elements.
 The part of muons that are transferred to muonic deuterium at natural
 deuterium concentration is small and has little impact on the process
 of muon transfer to higher-Z nuclei \cite{transfer-rate-to-D}; nuclear
 fusion therefore can also be neglected.
 The muons that are captured in muonic molecules drop out of the
 sequence of processes leading to transfer to higher-Z gases; since
 muonic molecule formation \cite{faifman} is much slower that muon
 transfer even at low admixture gas concentration (see
 Sect.~\ref{sect:transfer}) it will not be taken into account.
 As for the muons directly captured in muonic atoms of the higher-Z
 admixture gas, their fast cascade de-excitation produces prompt X-rays
 that can be clearly distinguished from the delayed X-rays that follow
 the transfer of the muon from $\mu^-p$.
 With all this in mind, we restrict our consideration to the following
 subset of processes:
 \begin{enumerate}
 \item
   thermalization of the muonic hydrogen atoms via elastic scattering of
   $\mu^-p$ by H$_2$ molecules;
 \item
   depolarization of the muonic hydrogen atoms in spin-flip scattering
   by H$_2$ molecules;
 \item
   muon transfer from the hydrogen to higher-Z muonic atom;
 \item
   disappearance of the muon by either a muon decay or nuclear muon
   capture.
 \end{enumerate}

 \subsection{Thermalization and depolarization of the muonic hydrogen
   atoms}

 Part of the energy released in the cascade de-excitation of $\mu^-p$
 atoms is transformed into kinetic energy of the latter, and the initial
 energy distribution in the ground $1s$ state is spread over a broad
 interval up to the keV range \cite{jens02,pohl06,faif08}, with the para
 ($F=0$) and ortho ($F=1$) spin states populated statistically.
 A phenomenological distribution of the initial kinetic energy that has
 given good results in simulations of preceding experiments is the sum
 of two Maxwellian distribution density, one corresponding to the target
 gas temperature, and another one with mean energy of 20~eV; it has been
 used in the present work too.
 Thermalization and depolarization occur in elastic and spin-flip
 scattering with the surrounding hydrogen and higher-Z atoms and
 molecules. The energy loss in collisions with the light H$_2$ is the
 main mechanism of thermalization, so that the rate of thermalization is
 little sensitive to the admixture concentration and depends only on the
 hydrogen density $\phi$ (or pressure $P$) and --- through the molecular
 cross sections --- on the temperature $T$.
 The same holds for the rate of depolarization.

\begin{figure}[htb]
   \centering
   \includegraphics[width=0.9\textwidth]{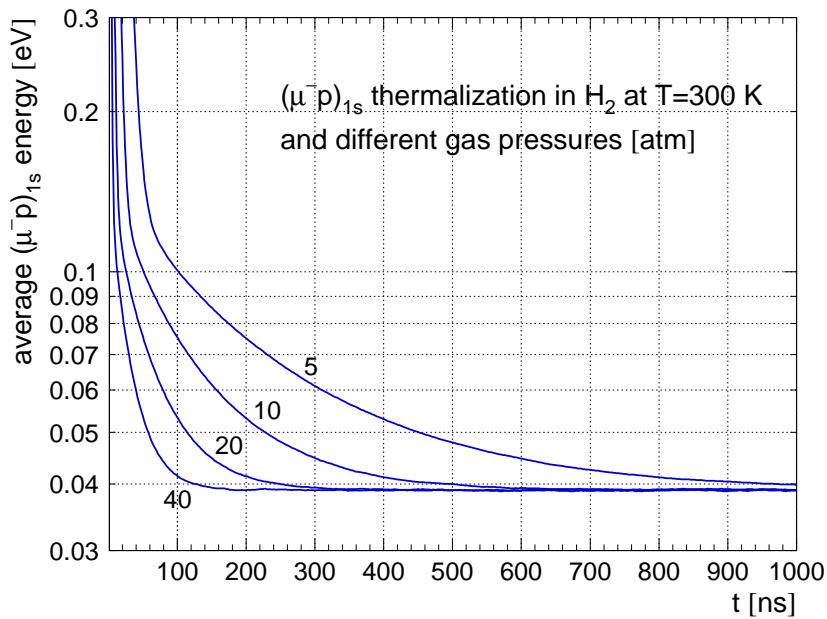}
   \caption{The average $(\mu^-p)_{1s}$ energy $\overline{E}(t;P,T)$
     versus time~$t$, for pressures $P=$5, 10, 20, and 40 atm
     (0.51, 1.01, 2.03, and 4.05 MPa).}
   \label{fig:eav_dens}
  \end{figure}

\begin{figure}[htb]
    \centering
    \includegraphics[width=0.9\textwidth]{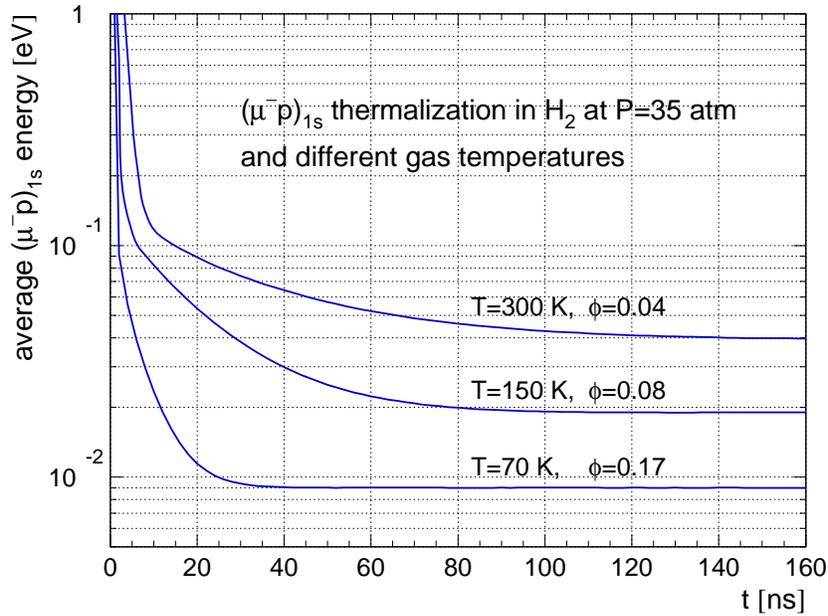}
    \caption{Time evolution of the average $(\mu^-p)_{1s}$ energy
      $\overline{E}(t;P,T)$ at a fixed pressure $P=35$~atm (3.55 MPa), for a set of
      temperatures $T$. $\phi$ is the $H_2$ gas density in units of LHD.}
    \label{fig:eav_temp_phi}
  \end{figure}

\begin{figure}[htb]
    \centering
    \includegraphics[width=0.9\textwidth]{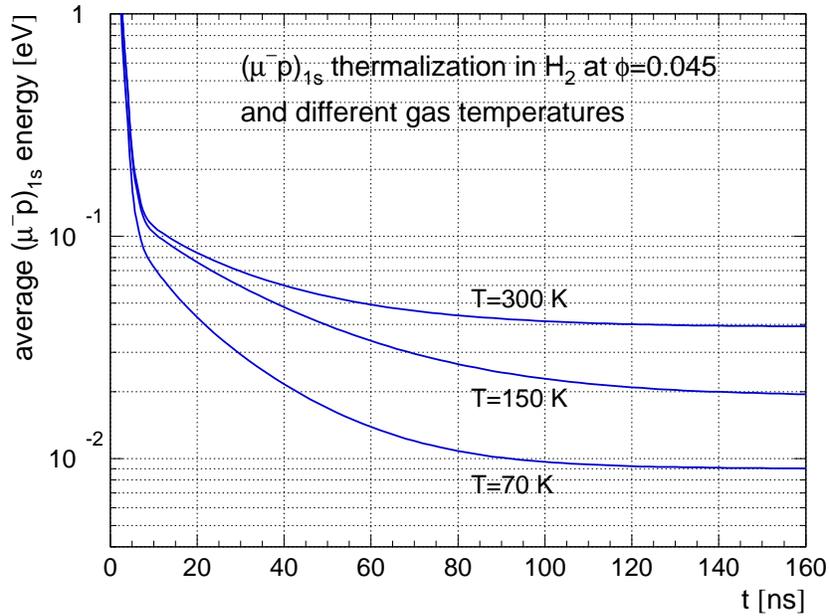}
    \caption{Time evolution of the mean $(\mu^-p)_{1s}$ energy
      $\overline{E}(t;\phi,T)$ at a fixed density $\phi=0.45$ in units of
      LHD, for a set of temperatures $T$.}
    \label{fig:eav_temp_pres}
  \end{figure}

 The process of thermalization is best illustrated with the time
 evolution of the average kinetic energy of $(\mu^-p)_{1s}$,
 $\overline{E}(t;P,T)$. Figure~\ref{fig:eav_dens} presents the Monte
 Carlo simulations for $\overline{E}(t;P,T)$ in pure hydrogen for a set
 of relevant pressures~$P$, at $T=300$~K.
  The plot clearly shows that the time needed for the average energy
  $\overline{E}(t;P,T)$ to approach the equilibrium energy is
  approximately inverse proportional to the pressure (i.e., density):
  $\overline{E}(t;\,kP,\,T)\sim\overline{E}(t/k;P,\,T), k>0$.
  In particular, at room temperature and 20 atm, the $(\mu^-p)_{1s}$
  atoms are ``practically'' thermalized already at $t=300$ ns. This
  conclusion is confirmed by the plots in Fig.~\ref{fig:eav_temp_phi} of
  $\overline{E}(t;P,T)$ at fixed pressure, for a set of values of the
  temperature and the corresponding hydrogen density~$\phi$, in the
  units of liquid hydrogen density (LHD).
  When the density is fixed, the thermalization time is practically the
  same for all discussed temperatures, which is shown
  in~Fig.~\ref{fig:eav_temp_pres}. At $\phi=0.045$, the $(\mu^-p)_{1s}$
  atoms are thermalized after about 150~ns, for all the temperatures.
  For times 5~ns~$\lesssim{}t\lesssim{}150$~ns, the time evolution of
  energy differs due to the different thermal energies of H$_2$
  molecules.  Thus, the gas density is the parameter that determines the
  thermalization time at small admixtures of higher-$Z$ gases. The same
  conclusion can be drawn for the time of quenching the $F=1$ state,
  since the collisions of $(\mu^-p)_{1s}$ atoms with H$_2$ molecules
  establish an effective mechanism of the downwards spin flip.

\begin{figure}[htb]
    \centering
    \includegraphics[width=0.9\textwidth]{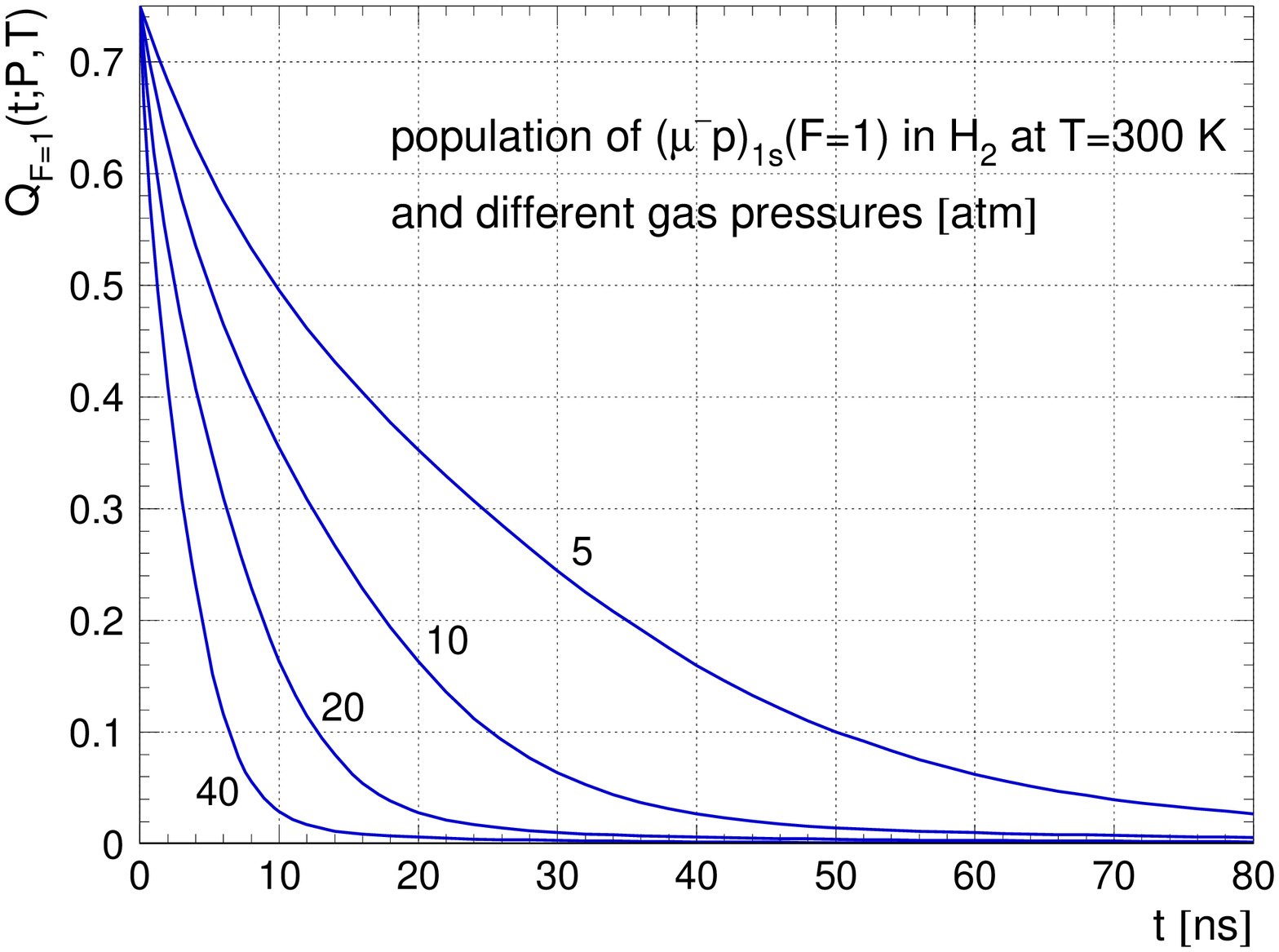}
    \caption{Time evolution of the population $Q_{F=1}(t; P,T)$ of the
      $F=1$ state in hydrogen at room temperature $T=300$ K, for a set
      of pressures $P=$5, 10, 20, and 40 atm (0.51, 1.01, 2.03, and 4.05
      MPa).}
    \label{fig:x_ogygen_01_O}
  \end{figure}

  The $(\mu^-p)_{1s}$ atom depolarization (i.e., the de-population of
  the ortho $F=1$ spin state) proves to be much faster than the
  thermalization process: the plot in Fig.~\ref{fig:x_ogygen_01_O} of
  the time evolution of $Q_{F=1}(t; P,T)$, the part of all
  $(\mu^-p)_{1s}$ atoms that occupy the ortho $F=1$ hyperfine level of
  the ground state (normalized by $Q_{F=1}(0; P,T)=3/4$), vanishes ---
  at the same temperature and pressure as in Fig.~\ref{fig:eav_dens} ---
  an order of magnitude faster than the atoms are thermalized.
  By summarizing the results of the above simulations, we conclude that
  to a good accuracy one can assume that the muonic hydrogen atoms are
  completely thermalized and depolarized after $t_0$ nanoseconds, where
  the rough estimate of $t_0$ reads:
  \begin{equation}
    t_0[\mathrm{ns}] \sim 20\,\times \,
    T[\mathrm{K}]\,/\,P[\mathrm{atm}] \,.
    \label{eq:thtime}
  \end{equation}

 \subsection{Muon decay and nuclear muon capture}
 \label{sect:decay}

 The proposed experimental method consists in measurements of the rate
 of muon transfer from thermalized $(\mu^-p)_{1s}$ atoms at time
 $t>t_0$; the muons that have decayed or been captured by this time are
 lost. The part of ``lost muons'' is $(1-\exp(-\lambda^*t_0))/\lambda^*$,
 $\lambda^*=\lambda_\mathrm{dec}+\lambda_\mathrm{cap}$ where
 $\lambda_\mathrm{dec}=0.45$~$\mu\mathrm{s}^{-1}$ and
 $\lambda_\mathrm{cap}=0.7$~ms$^{-1}$ are the free muon decay rate and
 the nuclear muon capture rate in $(\mu^-p)_{1s}$, both of them ---
 independent of hydrogen gas density or temperature. For $t_0=500$~ns,
 the losses are $\sim{}20$\% and increase with $t_0$.

 \subsection{Muon transfer to higher-Z muonic atoms}
 \label{sect:transfer}

 In the experiment for the measurement of the energy dependence of the
 rate of muon transfer to a higher-Z admixture nucleus we have the
 following parameters at our disposal to control the processes: the
 pressure and temperature of the target, the start-of-counting time
 $t_0$ and the concentration $c$ of the admixture, defined as
 $c=n_Z/(n_H+n_Z)$, where $n_H$ and $n_Z$ are the number concentrations,
 i.e. the number of species (atoms or molecules) of hydrogen and
 higher-Z element per cubic centimeter. The choice of optimal values of
 these parameters should guarantee that the maximal number of muon
 transfer events take place from thermalized $(\mu^-p)_{1s}$ atoms.

 The Monte Carlo simulations may have, in principle, only restricted
 validity because for most of the possible admixture gases only the
 average muon transfer rate is known. Oxygen is one of the very few
 elements for which there exist experimental and theoretical data about
 the energy dependence of the transfer rate. We chose to present the
 results of the simulations for a mixture of hydrogen and oxygen for
 this reason, and also because it will be the primary target of the
 oncoming experimental investigations.

\begin{figure}[htb]
  \centering
  \includegraphics[width=0.9\linewidth]{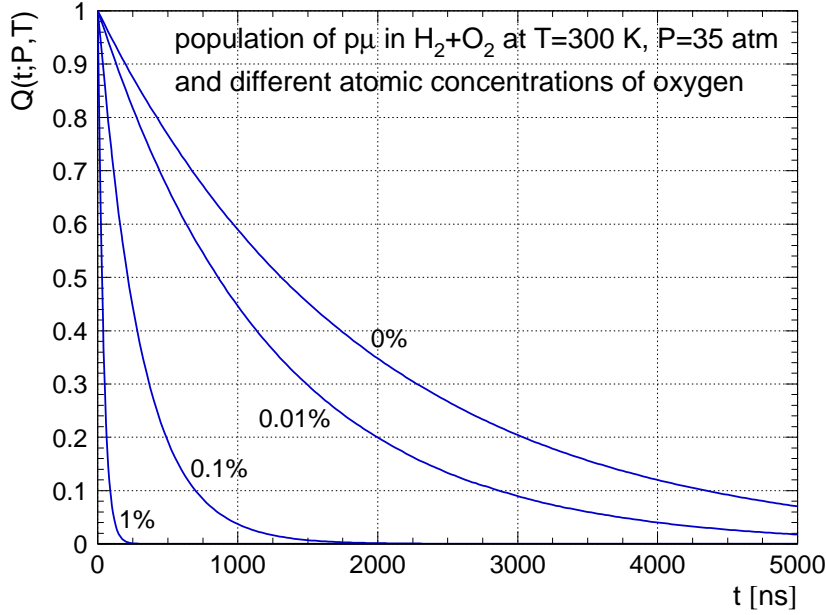}
  \caption{Number $Q(t;P,T)$ of the $(\mu^-p)_{1s}$ that have survived in
    H$_2$+O$_2$ mixture at $T=300$~K, $P=35$~atm (3.55 MPa), and oxygen
    concentrations $c=0$\%, 0.01\%, 0.1\%, and~1\% up to time $t$.
    $Q(t;P,T)$ is normalized to the number of thermalized depolarized
    $(\mu^-p)_{1s}$ atoms at time $t_0$. (Here for simplicity we have
    set $t_0=0$).}
  \label{fig:pop_all_conc}
 \end{figure}

\begin{figure}[htb]
  \centering
  \includegraphics[width=0.9\linewidth]{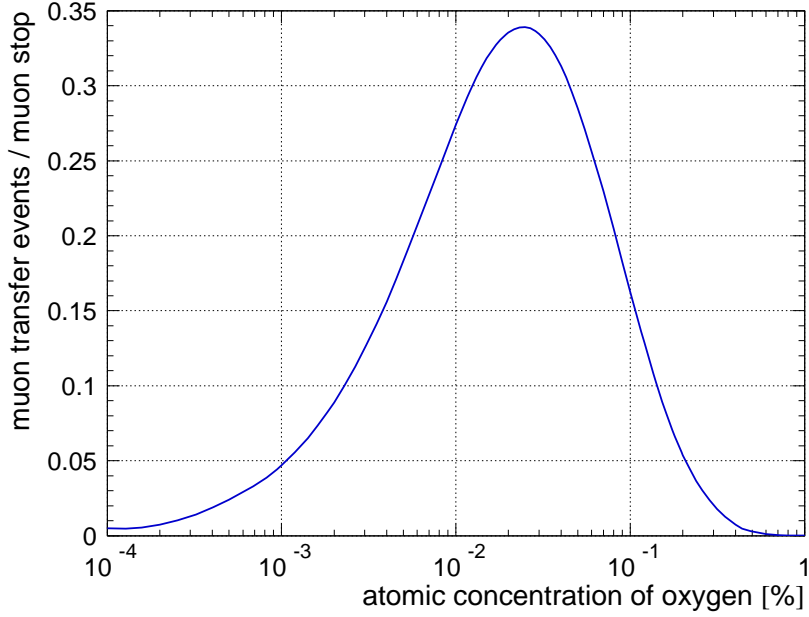}
  \caption{The number of muon transfer events from thermalized
    $(\mu^-p)_{1s}$ in H$_2$ at $T=300$~K and $P=35$~atm, normalized to
    the number of thermalized and depolarized muonic hydrogen atoms at
    $t=t_0=0$ versus the number concentration $c$ of oxygen.}
  \label{fig:xevent_c}
 \end{figure}

 In the experiment, it is crucial to have a~large number $(\mu^-p)_{1s}$
 atoms at times much larger than the time width of the prompt peak. In
 order to study this problem, the Monte Carlo simulations have been
 performed for a~fixed temperature and pressure of H$_2$+O$_2$ mixture
 and various oxygen concentrations. Figure~\ref{fig:pop_all_conc} shows
 the time evolution of the
 relative population of $(\mu^-p)_{1s}$ atoms at $T=300$~K and $P=35$~atm.
 The curve for $c=0$\%, i.e. for pure~H$_2$, determines the absolute
 upper limit on the number of surviving $(\mu^-p)_{1s}$ determined by
 the decay and capture rate $\lambda^*$. Any admixture of oxygen
 decreases this number. At $c\gtrsim{}1$\%, the hydrogen muonic atoms
 disappear before thermalization. Simulations with the FLUKA code
 \cite{fluka} for various oxygen concentrations at $P=35$~atm and
 $T=300$~K have shown that the fraction of muon stops in target
 practically does not depend on the concentration if $c\lesssim{}1$\%.
 The optimal oxygen concentration is the one that provides a~maximal
 number of muon transfer events from {\em thermalized} $(\mu^-p)_{1s}$
 out of a fixed number of muon stops in~H$_2$. The results of the Monte
 Carlo simulations for various oxygen concentrations at fixed
 temperature and pressure, shown in~Fig.~\ref{fig:xevent_c}, clearly
 demonstrate that the statistical uncertainty of the experimental data
 on the muon transfer rate --- which is inverse proportional to the
 squared root of the transfer events --- may be drastically reduced by
 careful planning of the experiment.  Our Monte Carlo prediction of the
 sharp dependence of the number of muon transfer events on the
 concentration of oxygen, which should be recalculated for any other
 target temperature and gas admixture, is a central result of the
 present paper.

 \section{Determining the energy dependence of the muon transfer rate
   from experimental data}
 \label{sect:rate}

 In this last section we present the mathematical model of the proposed
 experiment and derive the equations that allow for determining the
 energy dependence of the muon transfer rate from the experimental data
 on the temperature dependence of the muon transfer rate in thermalized
 mixture of hydrogen and a higher-Z gas. We consider the general case
 when the rate $\lambda$ of muon transfer in collisions of the muonic
 hydrogen atom with an atom of the heavier admixture does depend on the
 center-of-mass (CM) collision energy $\varepsilon$ of the species:
 $\lambda=\lambda(\varepsilon)$. The observable rate of muon transfer
 $\Lambda_{o}$ is the average of $\lambda$ over the thermal distribution
 of the collision energy $\rho(\varepsilon)$:
 \begin{equation*}
   \Lambda_{o}=\int \lambda(\varepsilon) \rho(\varepsilon)\,
   d\varepsilon \,.
 \end{equation*}
 In thermal equilibrium the distribution $\rho(\varepsilon)$ is
 nothing but the Maxwell-Boltzmann distribution:
 \begin{equation*}
   \begin{split}
     &
     \rho(\varepsilon)=\rho_{\rm MB}(\varepsilon;T)=
     \frac{1}{\varepsilon_T}\rho_0(\varepsilon/\varepsilon_T) \,,
     \\
     &
     \rho_0(x)=\frac{2}{\sqrt{\pi}}\sqrt{x}\exp(-x)\,, \quad
     \varepsilon_T=k_B T \,,
   \end{split}
 \end{equation*}
 where $T$ is the temperature and $k_B$ is Boltzmann's constant.

\begin{figure}[h]
   \centering
   \includegraphics[width=0.9\linewidth]{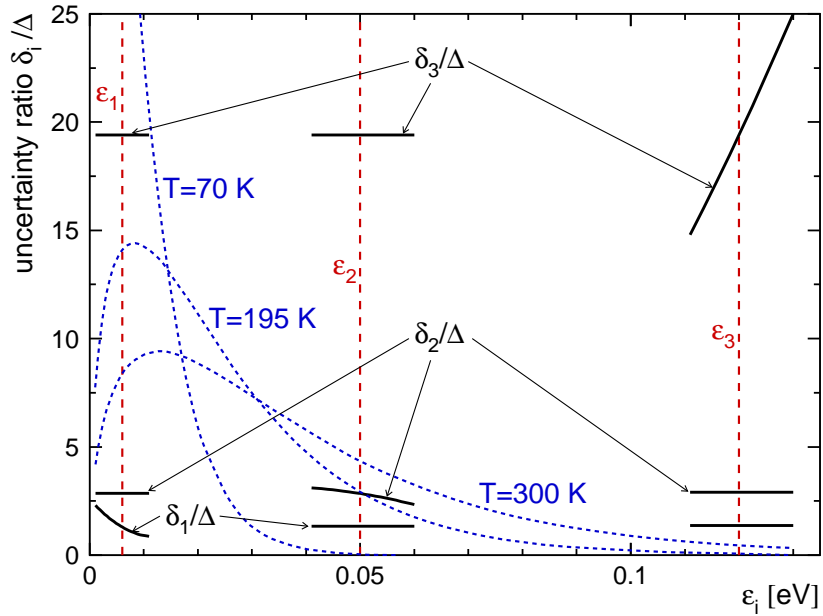}
   \caption{Dependence of the uncertainty ratios $\delta_i/\Delta$ on
     the reference energies $\varepsilon_j$, $j=1,2,3$.
     The thick solid piecewise curves display the response of
     $\delta_i/\Delta$ to variations of $\varepsilon_j$
     in the neighborhood of the value denoted by a dashed vertical
     line while keeping fixed the other two reference energies
     $\varepsilon_k, k\ne j$. The dotted curves are the
     Maxwell-Boltzman energy distribution densities for
     temperatures $T=70$ K, 195 K, and 300 K.}
   \label{fig:e3}
 \end{figure}

 Of primary importance is the behavior of $\lambda(\varepsilon)$ in the
 thermal and near epithermal energy range since the efficiency of the
 adopted experimental method for the measurement of the $\mu^-p$
 hyperfine splitting depends on the variations of $\lambda(\varepsilon)$
 in this range \cite{hfi2001}.  We use a polynomial approximation for
 $\lambda(\varepsilon)$:
 \begin{equation}
   \lambda(\varepsilon)=\sum\limits_{i=1}^{N_a}
   \lambda_i \prod\limits_{j=1, j\ne i}^{N_a}
   \frac{\varepsilon-\varepsilon_j}{\varepsilon_i-\varepsilon_j}
   \,, \quad
   \lambda_i=\lambda(\varepsilon_i) \,,
   \label{interp}
 \end{equation}
 where the values $\varepsilon_i, i=1,\ldots,N_a$ will be referred
 to as ``reference energies''.
 As long as the transfer rate $\lambda(\varepsilon)$
 is not expected to have any anomalous
 behavior outside the above range (e.g., abrupt decrease or growth at
 higher energies) we evaluate the integral for the observable muon
 transfer rate at temperature $T$ with a Gauss-type quadrature of
 appropriate rank $N_G$:
 \begin{equation}
   \Lambda(T)=\int \lambda(\varepsilon) \rho(\varepsilon;T) \,
   d\varepsilon= \int \rho_0(x) \lambda(\varepsilon_T\,x)\,dx=
 \sum\limits_{n=1}^{N_G} w_n \lambda(\varepsilon_T\,x_n) \,,
 \label{gauss}
 \end{equation}
 where $w_n$ and $x_n$ are the weights and nodes of the Gauss
 quadrature formula with weight function $\rho_0(x)$.
 Eqs. (\ref{interp},\ref{gauss}) establish a linear relation
 between $\Lambda(T)$ and the values $\lambda_i$:
 \begin{equation}
   \Lambda(T)=\sum\limits_{n=1}^{N_G}w_n
   \sum\limits_{i=1}^{N_a}
   \prod\limits_{j=1, j\ne i}^{N_a}
   \frac{k_B T\,x_n-\varepsilon_j}{\varepsilon_i-\varepsilon_j}\,
   \lambda_i \,.
   \label{lin0}
 \end{equation}
 This way, by measuring the values of the observable muon transfer rate
 $\Lambda_k=\Lambda(T_k)$ in thermalized target gas
 at temperatures $T_k, k=1,\ldots,N_a$, one
 will get a system of $N_a$ linear equations
 \begin{equation}
   \Lambda_k=\sum\limits_{i=1}^{N_a}M_{ki}\lambda_i \,, \quad
   M_{ki}=\sum\limits_{n=1}^{N_G}w_n\prod\limits_{j\ne i}
   \frac{k_B T_k\,x_n-\varepsilon_j}{\varepsilon_i-\varepsilon_j}
   \,.
   \label{mat}
 \end{equation}
 These equations can be resolved for the values
 $\lambda_i$ of the transfer rate
 at the reference energies:
 %
 $
   \lambda_i=\sum\limits_{k=1}^{N_a}M^{-1}_{ik}\Lambda_k \,,
 $
 %
 and thus the parameters in the polynomial
 approximation for $\lambda(\varepsilon)$ of Eq.~(\ref{interp})
 are expressed in terms of the experimental
 data $\{\Lambda_k\}$:
 \begin{equation}
   \lambda(\varepsilon)=\sum\limits_{i=1}^{N_a}
   \prod\limits_{j=1, j\ne i}^{N_a}
   \frac{\varepsilon-\varepsilon_j}{\varepsilon_i-\varepsilon_j}.
   \sum\limits_{k=1}^{N_a}M^{-1}_{ik}\Lambda_k
   \label{interp1}
 \end{equation}
 While these transformations are quite straightforward, it is worth
 briefly discussing the uncertainty of the parameters
 $\lambda_i=\lambda(\varepsilon_i)$, induced by the experimental
 uncertainty of $\Lambda_k=\Lambda(T_k)$. If assuming that $\Lambda_k$
 are normally distributed around $\bar{\Lambda}_k$ with standard
 deviation $\Delta_k$, the parameters $\lambda_i$ will be normally
 distributed with standard deviation $\delta_i$, where
 \begin{equation}
   \delta_i^2=
   \sum\limits_{k=1}^{N_a}\left(M_{ik}^{-1}\right)^2\Delta_k^2 \,.
   \label{eq:uncert}
 \end{equation}
 %
 The specific values of $\delta_i$ depend on the choice of target
 temperatures ($T_k$, $k=1,\ldots{},N_a$) (that may be predetermined by
 the available cooling devices) and of the reference energies
 $\varepsilon_i$ (whose optimal choice will be done after simulations of
 the muonic hydrogen HFS experiment). Consider as an example a series of
 three measurements at 70~K (cooling with liquid nitrogen), 195~K
 (liquid carbon dioxide) and 300~K (room temperature), and the set of
 reference energies $\varepsilon_1=0.006$~eV, $\varepsilon_2=0.05$~eV,
 and $\varepsilon_3=0.12$~eV. For simplicity, we assume that all
 experimental uncertainties are equal: $\Delta_k=\Delta$,
 $k=1,\ldots,3$. The accuracy of the proposed method is assessed with
 the ratios $\delta_i/\Delta$ that show how much the uncertainties of
 the calculated values $\lambda_i$ exceed the uncertainty of the
 experimental data $\Lambda_k$.
 On Fig.~\ref{fig:e3} we plot (with thick solid curves) the response of
 $\delta_i/\Delta$ to the variations of each of the three reference
 energies $\varepsilon_j,j=$1,2,3, in the neighborhood of the values listed above
 (denoted with dashed vertical lines) while keeping the other two fixed.
 We see that:
 \begin{itemize}
   \item the uncertainty of $\lambda(\varepsilon)$ increases very
   fast with the reference energy $\varepsilon$.
  This is not surprising: the contribution to the observable
  transfer rate $\Lambda(T)$ from the ``code'' of the
  Maxwell-Boltzman distribution $\rho_{\rm MB}(\varepsilon;T)$
  (plotted with dotted lines for the selected temperatures $T_k$)
  decreases exponentially with energy;

  \item the uncertainty $\delta_i$ is little sensitive to
  variations of the reference energies $\varepsilon_j, j\ne i$.
 \end{itemize}

 Eq.~\ref{eq:uncert} allows for estimating the accuracy goals
 $\Delta_k$ in the
 measurements of $\Lambda_k$ that will lead to uncertainties
 $\delta_i$ of the
 calculated muon transfer rates $\lambda_i$ within the limits, required
 for planning and optimizing the muonic hydrogen HFS experiment
 \cite{hfi2001}.
 The example discussed above shows that,
 with a relatively modest statistics of $\sim10^6$
 muon transfer events,
 the transfer rate $\lambda(\varepsilon)$ can be determined
 in the whole range of epithermal energies of interest
 with uncertainty of the order of 1\%
 -- far below the uncertainty of the currently
 available data about the energy dependence of the muon transfer
 rates.

 The numerical results are independent of the rank $N_G$ of the Gauss
 quadrature formulae provided that $N_a\le 2N_g$. The nodes and weights
 of two lower rank quadratures are given in Table \ref{table:gauss}.
 \begin{table}[htb]
   \caption{Examples of quadrature formula for averaging over the
     Maxwell-Boltzmann energy distribution}
   \label{table:gauss}
   \begin{tabular}{l|ll}
     \hline\hline
     $N_g$ & \multicolumn{1}{c}{Nodes $x_n$} & \multicolumn{1}{c}
     {Weights $w_n$} \\ \hline
     & \phantom{1}0.666326  & 0.6400012 \\
     3 & \phantom{1}2.800775  & 0.3445751 \\
     & \phantom{1}7.032899 & 0.0154237 \\
     \hline
     & \phantom{1}0.431399 & 0.4180087 \\
     & \phantom{1}1.759754 & 0.4655516 \\
     5 & \phantom{1}4.104465 & 0.1103327 \\
     & \phantom{1}7.746704 & 0.00606325 \\
     & 13.45768 & 0.00004372 \\ \hline
   \end{tabular}
 \end{table}
 %
 %

 \section{Conclusions}

 Using a realistic model of the processes with muonic hydrogen atoms in
 the ground state that occur in mixture of hydrogen and a higher-Z gas,
 we have found by means of Monte Carlo simulations the experimental
 conditions (admixture concentration and time gate) for preselected
 pressure and temperature which optimize the efficiency of the
 measurement of the temperature dependence of the average rate of muon
 transfer from muonic hydrogen to the higher-Z admixture. We have also
 developed a computational technique allowing to determine from these
 data the energy dependence of the muon transfer rate and give an
 estimate of the experimental uncertainty of the latter. These results
 will be implemented in the methodology of the oncoming experiment FAMU
 \cite{famu}, to be performed in 2015 at the RIKEN-RAL muon facility.

 \section*{Acknowledgments}

 This work has been partly supported under the bilateral agreement of
 the Bulgarian Academy of sciences and the Polish Academy of Sciences.

\end{document}